\documentclass[twocolumn,showpacs,preprintnumbers,amsmath,amssymb]{revtex4}
\usepackage{graphicx}

\begin{document}
\title{Noncollinear magnetic order in Quasicrystals}

\author{E.Y.~Vedmedenko,$^{1}$ U.~Grimm,$^{2}$ and R.~Wiesendanger$^{1}$}

\affiliation{$^{1}$Institut f\"ur Angewandte Physik,
Jungiusstr. 11, 20355 Hamburg, Germany\\
$^{2}$Applied Mathematics Department,
The Open University, Walton Hall, Milton Keynes MK7 6AA, UK}

\begin{abstract}
Based on Monte-Carlo simulations, the stable magnetization
configurations of an antiferromagnet on a quasiperiodic tiling are
derived theoretically. The exchange coupling is assumed to decrease
exponentially with the distance between magnetic moments.  It is
demonstrated that the superposition of geometric frustration with the
quasiperiodic ordering leads to a three-dimensional noncollinear
antiferromagnetic spin structure. The structure can be divided into
several ordered interpenetrating magnetic supertilings of different
energy and characteristic wave vector.  The number and the symmetry of
subtilings depend on the quasiperiodic ordering of atoms.
\end{abstract}
\maketitle

The last few years have shown a boom in investigations of the spin
order in antiferromagnetic films \cite{Kurz,Heinze} motivated by the
dramatic changes in the magnetic properties of such systems induced by
frustration.  In contrast to the rather well studied spin structure of
antiferromagnets on periodic lattices, the antiferromagnetic ordering
of quasicrystals is subject of ongoing scientific debate. Whereas an
experimental finding of long-range antiferromagnetic order in
rare-earth icosahedral quasicrystals \cite{Charrier} turned out to be
an artefact \cite{Sato1}, theoretical models that deal with magnetism
on quasicrystals \cite{Wessel} are known to exhibit long-range
magnetic order.  Recent inelastic neutron scattering experiments on
the Zn-Mg-Ho icosahedral quasicrystal \cite{Sato2} revealed a very
peculiar diffuse scattering pattern with icosahedral symmetry at
temperatures below 6K. Such a pattern, in principle, can originate
from a noncollinear spin arrangement first suggested by Lifshitz from
pure geometrical considerations
\cite{Lifshitz,Lifshitz1,Lifshitz2}. However, real-space magnetic
configurations leading to those long wave vector correlations remain
obscure despite recent interesting results for quantum spins
\cite{Wessel}. Thus, the knowledge about the spin structure on
quasiperiodic tilings is of basic importance for experiments as well
as for theoretical predictions of new phenomena, which can be expected
due to nontrivial frustration effects \cite{Naumis}.

\begin{figure}
\includegraphics[width=.85\columnwidth]{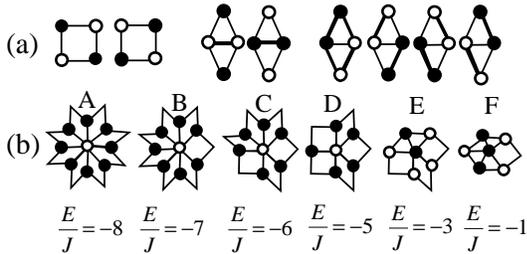}
\caption{Configurations for a frustrated Ising antiferromagnet on
(a) elementary tiles and (b) six local environments of the
Ammann-Beenker tiling.  Bold lines denote the frustrated bonds. The
open and filled circles represent different spins.
\label{fig01}}
\end{figure}

The patterns found in our theoretical study provide an explanation for
the origin of the antiferromagnetic modulations observed
experimentally in Ref.~\cite{Sato2}.  While the spin order in
antiferromagnets is usually characterized by a periodic modulation
described by wave vectors on the order of inverse atomic distances,
the spin order in antiferromagnetic quasicrystals admits
three-dimensional noncollinear structures consisting of several
interpenetrating subtilings with longer wave vectors. Here we report
on the details of the low-temperature antiferromagnetic ordering and
the map of the local frustration for the octagonal tiling.

We discuss the antiferromagnetic Hamiltonian
\begin{equation}\label{1}
H = J_{ij}\sum\limits_{\left\langle {i,j} \right\rangle}
\mathbf{S}_{i} \cdot \mathbf{S}_{j} -
K_{1}\sum\limits_{i} (\mathbf{S}_{i}^{z})^2
\end{equation}
where $\mathbf{S}_{i}$ is a three- or two-dimensional unit vector in
the case of classical vector or $xy$-spins, and $\mathbf{S}_{i}^{z}$
is equal to $\pm 1$ in the case of Ising spins (so
$\mathbf{S}_{i}^{x}=\mathbf{S}_{i}^{y}=0$); $\left\langle {i,j}
\right\rangle$ denotes the nearest neighbor pairs.  For an
antiferromagnetic system, the exchange parameter $J_{ij}$ is positive,
and neighboring antiparallel spins contribute a lower energy than
parallel neighbors. The coefficient $K_1$ is the first-order
anisotropy constant. Our Monte-Carlo simulations have been carried out
on finite Ammann-Beenker tilings with free boundary conditions. The
procedure is a simulated annealing method with at least 15 successive
temperature steps \cite{Vedmedenko}. At each temperature, the
convergence of the relaxation process towards equilibrium has been
observed for any initial configuration after a few thousand Monte
Carlo steps per spin. Hence, the single-spin-update algorithm is
efficient in our case. At the end of the cooling down process, the
total energy is just fluctuating around its mean equilibrium value. To
reduce boundary effects only the core of a tiling has been
analyzed. The samples on the octagonal Ammann-Beenker structure, which
we shall concentrate on in what follows, are circular, containing
2193, 11664 and 53018 magnetic moments.

The octagonal tiling consists of two motifs: a square and a rhombus of
equal edge lengths $a$ (Fig.~\ref{fig01}(a)). The diagonal bonds are,
usually, neglected in the calculations \cite{Wessel,Grimm}. We find
this disregard physically questionable as the exchange coupling
increases exponentially with decreasing interatomic distance.  In the
present investigation, the short diagonal of the rhombus and the sides
of the motifs have been considered as nearest neighbors. We
distinguish the two cases $J_d>2J$ and $J_d<2J$, where $J_{d}$ denotes
the interaction along the short diagonal and the interaction strength
along the sides $J$ is unity. The first case corresponds to a rapid
growth of the exchange coupling with decreasing interatomic
distance. The two nearest-neighbor bonds form six local environments
with coordination numbers varying from 5 to 8 as shown in
Fig.~\ref{fig01}(b). They occur with relative frequencies $\nu^{}_{A}=
17-12\sqrt{2}\approx 2.9\%$, $\nu^{}_{B}= -41+29\sqrt{2}\approx
1.2\%$, $\nu^{}_{C}= 34-24\sqrt{2}\approx 5.9\%$, $\nu^{}_{D}=
-14+10\sqrt{2}\approx 14.2\%$, $\nu^{}_{E}= 6-4\sqrt{2}\approx
34.3\%$, and $\nu^{}_{F}= -1+\sqrt{2}\approx 41.4\%$
\cite{Baake}. Taking into account the short diagonals of the rhombic
tiles increases the average coordination number of the tiling from $4$
(the value without diagonals) to
$8\nu^{}_{A}+7\nu^{}_{B}+6\nu^{}_{C}+5(\nu^{}_{D}+\nu^{}_{E}+\nu^{}_{F})
=8-2\sqrt{2}\approx 5.17$.

\begin{figure}
\includegraphics[width=.85\columnwidth]{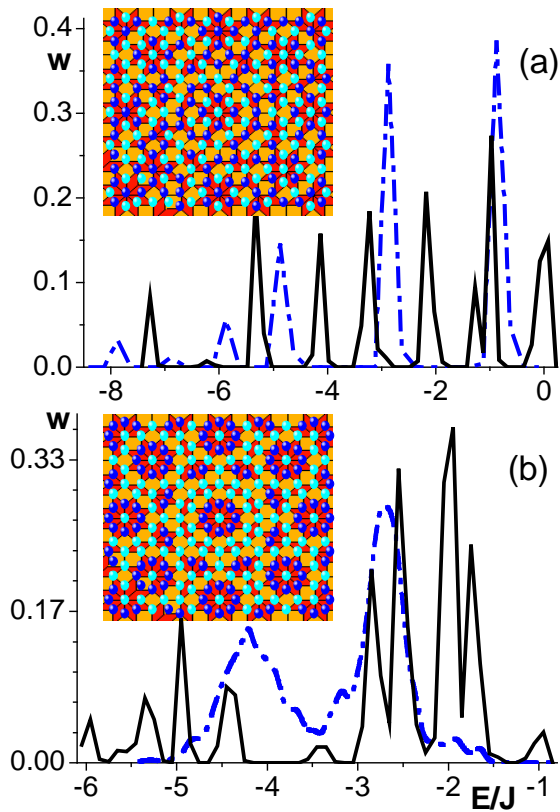}
\caption{The frequency distribution of the energy per spin on the
octagonal tiling for (a) Ising and (b) vector spins. Solid lines
correspond to the case $J_d<2J$, dashed lines to $J_d>2J$. Purely
antiferromagnetic interaction at $kT=0.01J$ is considered.
Top-views of portions of Monte-Carlo configurations with
underlying tilings are shown as insets. The light and dark circles
represent different spins in (a) and different energies in (b),
respectively. \label{fig02}}
\end{figure}

First we discuss the Ising system. The square tile of the octagonal
structure is non-frustrated as every pair of the moments can be chosen
to be antiparallel (Fig.~\ref{fig01}(a)). If we had not taken the
short diagonals of the rhombic tiles into account, the same would have
been true for the entire tiling, and there would be no frustration,
because the rhombic tiling is bipartite.  Now, we consider spins on
short diagonals as nearest neighbors, the rhombic tiles are always
frustrated.  If the energy of one nearest neighbor pair is minimized
by having antiparallel spins, the third and forth spins cannot be
chosen to minimize the energy of both of its neighbors
(Fig.~\ref{fig01}(a)). The magnetic moment will necessarily be
parallel to one of the neighbors. For $J_d<2J$ two out of six possible
configurations have smaller energy as they possess only one pair of
parallel nearest neighbors per rhombus instead of two
(Fig.~\ref{fig01}(a)). In this case spins can have one of six possible
energy values corresponding to different local environments
(Fig.~\ref{fig01}(b)). For $J_d>2J$ the four configurations with two
parallel bonds have lowest energy as their weight is smaller than that
of the strong diagonal coupling. The second case comprises much more
different possibilities of energy distribution. To give a quantitative
description of the local frustration we introduce a local parameter
$f$
\begin{equation}\label{2}
f = \frac{{|E_{id} | - |E_i |}}{{|E_{id} |}}
\end{equation}
where $E_i$ is an actual energy of a spin $i$ and $E_{id}$ is a ground
state energy of a relevant unfrustrated vertex.  With this
nomenclature, only the central spins of the vertices $F$ and $E$ are
magnetically frustrated $f_{F}=0.4$ and $f_{E}=0.8$ for
$J_d=J<2J$. The Monte-Carlo simulations confirm our reasoning based on
the analysis of frustration. Fig.~\ref{fig02}a gives the frequency
distribution of the exchange energy per atom $E$ for two cases and a
top-view of a portion of Ising configuration with $J_d>2J$. The energy
distribution for $J_d<2J$ simply reproduces the frequency of $6$
vertex configurations. The "up" and "down" configurations are
perfectly ordered and coincide with the black-and-white model of
Niizeki \cite{Niizeki}.  For large $J_d$ we find $8$ possible energy
values. The "up" and "down" subtilings, however, are spatially
disordered (see inset Fig.~\ref{fig02}a). We have calculated the
magnetic structure factor
\begin{equation}\label{3}
S^{zz}(\mathbf{k}) = \frac{1}{N}\sum\limits_{\mathbf{r},\mathbf{r}'}
e^{i\mathbf{k}\cdot(\mathbf{r}-\mathbf{r}')}
\left\langle S_{\mathbf{r}}^z S_{\mathbf{r}'}^z  \right\rangle
\end{equation}
using the Monte-Carlo data for different samples. Here $\mathbf{k}$ is
the wave vector and $S_{\mathbf{r}}^z$ is a vertical component of a
magnetic moment at the position ${\mathbf r}$. The diffraction pattern
of the Niizeki configuration coincides with that of quantum
Monte-Carlo calculations (Fig.~5c,d of Ref.~\cite{Wessel}) and
theoretical prediction \cite{Lifshitz2}, while the intensity map of
the configuration Fig.~\ref{fig02}a is almost structureless. It means
that Ising solution with $J_d<2J$ reproduces in essence the
antiferromagnetic superstructure, corresponding to a modulation vector
$\mathbf{q} = (\frac{1}{2},\frac{1}{2},\frac{1}{2},\frac{1}{2})_{a^*}$
\cite{Sato2} in the octagonal tiling, whereas stronger coupling leads
to a spin-glass state.

\begin{figure}
\includegraphics[width=.99\columnwidth]{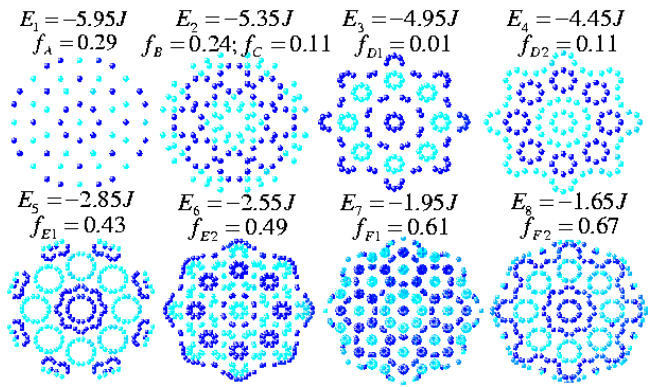}
\caption{(color on-line) Spatial distribution of magnetic moments
belonging to eight subtilings of a noncollinear configuration on
an octagonal tiling consisting of 2193 spins. $J_d>2J$. The
light and dark circles represent
positive and negative $x$-components of the magnetization. The
in-plane components are not given for the sake of simplicity.
Average values of the exchange energy $E$ and the local
frustration $f$ per spin are indicated. \label{fig03}}
\end{figure}

An exciting question is if the further minimization of the total
energy and frustration by means of the noncollinear alignment of
magnetic moments is possible.  At first glance the magnetic structure
of the low-temperature pure antiferromagnetic configuration seems to
be rather disordered. The analysis of the local energies, however,
reveals several characteristic energetic maxima in the frequency
distribution shown in Fig.~\ref{fig02}(b).  The simple existence of
the peaks means that there exist different sorts of magnetic moments
having well-defined relative orientation to their nearest
neighbors. This orientation, however, is not associated with any
absolute direction in space. Therefore, in accordance with the
Mermin-Wagner theorem \cite{Mermin}, no long-range order exists in
two-dimensions with continuous symmetry, because thermal fluctuations
result in a mean-square deviation of the spins from their equilibrium
positions which increases logarithmically with the size of the
system. The addition of a very weak anisotropy, which often exists in
real samples, does not change the distribution of the exchange energy,
but permits to anchor the absolute spatial orientation of the
magnetization. Nevertheless, at first glance the total structure still
looks spin-glass like. In the following, we will show that the
antiferromagnetic structure of the octagonal tiling is perfectly
ordered, but the order is non-trivial and unusual for periodic
crystals. We concentrate further description on 3D vector spins while
similar results for $xy$-spins have been obtained.

To obtain an absolute symmetry axis, we apply a very weak out-of-plane
anisotropy $K_1\approx 10^{-3}J$ to the system. The squared vertical
component of magnetization $(S^{z})^2$ becomes finite. The positions
of the energy peaks on the frequency diagram remain unchanged. All
maxima are different from those of the Ising model. It means that the
angles between the neighboring magnetic moments are not always equal
to $180^\circ$ or $0^\circ$, i.e., the magnetic structure is
noncollinear. The different number of peaks --- eight for $J_d<2J$ and
two for $J_d>2J$ (Fig.~\ref{fig02}(b)) --- already tells us that, in
contrast to the Ising case, the maxima do not coincide with the 6
vertices of the tiling. The minimal possible local energy increases
from $-8J$ to approximately $-6J$ for $J_d=J$ or $-5.44J$ for
$J_d=2.2J$. The average energy per spin, however, decreases by more
than $0.3J$ and reaches the value of $E\approx-2.85J$ and
$E\approx-3.30J$ respectively. Hence, the increase of the entropy
permits to minimize the average frustration and the total energy of
the system.

\begin{figure}
\includegraphics[width=.85\columnwidth]{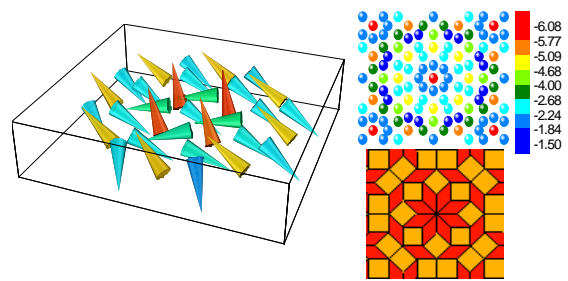}
\caption{Perspective view of a portion of a Monte-Carlo
configuration on an octagonal tiling.  The top view of the patch
and the energy map are shown as insets. Magnetic moments are
represented as cones. The cones are colored according to their
vertical magnetization, changing gradually from red for "up" to
blue for "down" spins. In the energy map inset, the colors encode
the energy per moment. \label{fig04}}
\end{figure}

Spatial arrangements of the magnetic moments as a function of the
exchange energy are given in Fig.~\ref{fig03} for $J_d<2J$ and in the
inset to Fig.~\ref{fig02}(b) for $J_d>2J$. Each configuration of
Fig.~\ref{fig03} represents a certain energy range corresponding to
one of the eight peaks in the spectrum of Fig.~\ref{fig02}(b). Colors
represent the $x$-projection of the magnetization. The magnetic
moments form 8 subtilings of different energy ($E_{1},\ldots,E_{8}$)
which generally do not coincide with a specific vertex type. The
splitting of the energy and frustration levels is described in detail
in Fig.~\ref{fig03}. For example the vertices $B$ and $C$
(Fig. \ref{fig01} belong to the same energy maxima $E_2$ but have
different local frustration $f_{B}=0.24$, $f_{C}=0.11$
(Fig.~\ref{fig03}).  At the same time the central spin of the vertex
$D$ can have either the energy $E_3$ or $E_4$ and, therefore, can have
two different values of the frustration $f_{D1}=0.01$ and
$f_{D2}=0.11$ depending on local surroundings. Thus, every
configuration of the Fig.~\ref{fig03} can enclose either a part of the
atomic places belonging to one vertex type or two different vertex
types together.  Nevertheless all structures have a perfect general
spatial ordering. Each subtiling can be separated into the
energetically degenerate `right' and `left' parts which also have a
perfect quasiperiodic arrangement. However, not all `right' or `left'
moments have identical orientation in space. Fig.~\ref{fig04} shows a
perspective view of a portion of typical Monte-Carlo configuration and
corresponding energy map. The central magnetic moment has the lowest
energy and belongs to the $E_1$ subtiling. Its 8 nearest neighbors
have identical energies and correspond to the energy $E_7$ despite
having different sets of mutual angles. The moments forming the next
ring have energy $E_6$.  The last ring consists of the alternating
$E_3$ and $E_6$ spins. Fig.~\ref{fig04} shows one of the radially
symmetric vertices. However, in the octagonal tiling vertices with
different surrounding can also be found. The energy distribution is
then different. Hence, the magnetic structure for $J_d<2J$ is
noncollinear and consists of eight interpenetrating subtilings. For
$J_d>2J$ we find only two subtilings of different energy.

\begin{figure}
\includegraphics[width=.99\columnwidth]{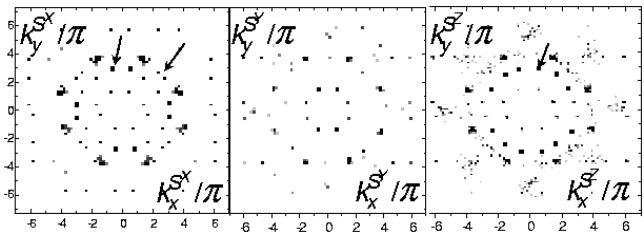}
\caption{The calculated Bragg scattering of $S^x$, $S^y$ and $S^z$
component of magnetization for the antiferromagnetic
superstructure. Reflexes indicated by arrows are new in comparison
to previous studies \cite{Wessel}.\label{fig05}}
\end{figure}

A frequency distribution of the angle between nearest neighboring
moments shows five characteristic angles close to $60^\circ$,
$80^\circ$, $120^\circ$, $140^\circ$ and $180^\circ$ for small $J_d$
and a single mutual angle of $110^\circ$ for large $J_d$.  Due to this
noncollinearity the energy of the system is decreased.  The
diffraction pattern of the whole structure is more complex than that
of the Ising or the quantum-mechanical \cite{Wessel} model. As the
spin structure is noncollinear, not only the structure factor
$S^{zz}$, but also $S^{xx}$ and $S^{yy}$ can be recognized (see
Fig.~\ref{fig05}). The eightfold $S^{xx}$ and $S^{zz}$ patterns
contain additional long wave-vector peaks which could not be
identified in the previous investigations \cite{Wessel}. In dependence
on the anisotropy (or on the initial random configuration for $K_1=0$)
new peaks also occur in $S^{yy}$. The Bragg reflexes found in our
study select a subset of the wave vectors given in
Ref.~\cite{Lifshitz2} where $n_1+n_2+n_3+n_4$ is odd. Peaks with
$n_1+n_2+n_3+n_4$ even are extinct. According to the nomenclature of
Ref.~\cite{Lifshitz2}, the following wave vectors can be identified:
$(1,0,0,0)$, $(1,-1,1,0)$, $(3,-2,1,1)$, $(3,-1,-1,2)$, $(1,1,-1,0)$,
$(1,0,1,-1)$, $(0,2,-1,0)$, $(0,0,1,-2)$, $(-1,0,1,-3)$, $(0,2,-2,1)$,
$(0,1,-2,2)$. Hence, the noncollinearity of the spin structure gives
rise to selection rules different from those of collinear models
\cite{Lifshitz, Wessel}. With increasing sample size the peaks become
more diffuse and may correspond to the diffuse scattering signal of
Ref.~\cite{Sato2}.

In conclusion, we demonstrate that the frustrated classical Ising
system with antiferromagnetic coupling on a quasiperiodic octagonal
tiling is perfectly ordered. All spins can be divided into 6
quasiperiodic (in the 3D physical space) or 6 periodic (in 6D periodic
crystal) subtilings of different energy.  Each subtiling corresponds
to the one of 6 vertex types of the Ammann-Beenker structure and is
degenerated for `up' and `down' magnetic moments. Quantitatively, only
two out of six subtilings are frustrated with the local coefficients
$f_{E}=0.4$ and $f_{F}=0.8$. The vector spin system admits a
three-dimensional noncollinear magnetic structure.  For $J_{d}<2J$,
the whole structure can be decomposed into 8 subtilings of different
energy which generally do not coincide with a specific vertex
type. All subtilings are frustrated. However, the total degree of
frustration and the energy of the system is minimized compared to the
noncollinear case.  The subtilings are degenerated with respect to the
spin direction. The codirectional spins of every subtiling reveal
perfect quasiperiodic ordering with a wave vector which is specific
for a given subtiling.

We thank R. Lifshitz for helpful comments and indexing of the
Bragg peaks. Financial support from the Interdisciplinary
Nanoscience Center Hamburg (INCH) is gratefully acknowledged.

\end{document}